\documentclass[twocolumn]{aastex63}

\usepackage{array,multirow}
\usepackage{amsmath}
\usepackage{wrapfig}
\usepackage{acronym}
\usepackage{enumitem}
\usepackage{lineno}
\usepackage{graphicx}
\usepackage{comment}

\graphicspath{{./}{figures/}}

\acrodef{AGN}{active galactic nuclei}
\acrodef{BH}{black hole}
\acrodefplural{BH}[BHs]{black holes}
\acrodef{BPS}{binary population synthesis}
\acrodef{BNS}{binary neutron star}
\acrodef{CCSN}{core-collapse supernova}
\acrodefplural{CCSN}[CCSNe]{core-collapse supernovae}
\acrodef{CC}{core-collapse}
\acrodef{CE}{common-envelope}
\acrodef{CO}{compact object}
\acrodef{C--O}{carbon-oxygen}
\acrodef{DCO}{double compact object}
\acrodef{ECSN}{electron-capture supernova}
\acrodefplural{ECSN}[ECSNe]{electron-capture supernovae}
\acrodef{F12d}{Fryer--12--delayed}
\acrodef{GC}{globular cluster}
\acrodef{GW}{gravitational wave}
\acrodefplural{GW}[GWs]{gravitational waves}
\acrodef{GWTC--3}{Gravitational-Wave Transient Catalog}
\acrodefplural{GC}[GCs]{globular clusters}
\acrodef{H}{hydrogen}
\acrodef{HMS--HMS}{hydrogen main-sequence}
\acrodef{He}{helium}
\acrodef{IMF}{initial mass function}
\acrodef{MS}{main-sequence}
\acrodef{MT}{mass transfer}
\acrodef{MSP}{millisecond pulsar}
\acrodefplural{MSP}[MSPs]{millisecond pulsars}
\acrodef{MW}{Milky Way}
\acrodef{NN}{nearest-neighbor}
\acrodef{NS}{neutron star}
\acrodefplural{NS}[NSs]{neutron stars}
\acrodef{RLO}{Roche-lobe overflow}
\acrodef{S16}{Sukhbold--16}
\acrodef{SN}{supernova}
\acrodefplural{SN}[SNe]{supernovae}
\acrodef{USSN}{ultra-stripped supernova}
\acrodefplural{USSN}[USSNe]{ultra-stripped supernovae}
\acrodef{ZAMS}{zero-age main-sequence}

\submitjournal{ApJ}

\begin{document}
\title{Challenges in Forming Millisecond Pulsar--Black Holes from Isolated Binaries}

\author[0000-0002-8883-3351]{Camille\, Liotine}
\affiliation{Department of Physics and Astronomy, Northwestern University, 2145 Sheridan Road, Evanston, IL 60208, USA}
\affiliation{Center for Interdisciplinary Exploration and Research in Astrophysics (CIERA), Northwestern University, 1800 Sherman Ave, Evanston, IL 60201, USA}

\author[0000-0001-9236-5469]{Vicky\, Kalogera}
\affiliation{Department of Physics and Astronomy, Northwestern University, 2145 Sheridan Road, Evanston, IL 60208, USA}
\affiliation{Center for Interdisciplinary Exploration and Research in Astrophysics (CIERA), Northwestern University, 1800 Sherman Ave, Evanston, IL 60201, USA}
\affiliation{NSF-Simons AI Institute for the Sky (SkAI),172 E. Chestnut St., Chicago, IL 60611, USA}

\author[0000-0001-5261-3923]{Jeff\, J.\,Andrews}
\affiliation{Department of Physics, University of Florida, 2001 Museum Rd, Gainesville, FL 32611, USA}
\affiliation{Institute for Fundamental Theory, 2001 Museum Rd, Gainesville, FL 32611, USA}

\author[0000-0002-3439-0321]{Simone\,S.\,Bavera}
\affiliation{Département d’Astronomie, Université de Genève, Chemin Pegasi 51, CH-1290 Versoix, Switzerland}
\affiliation{Gravitational Wave Science Center (GWSC), Université de Genève, CH1211 Geneva, Switzerland}

\author[0000-0002-6842-3021]{Max\,Briel}
\affiliation{Département d’Astronomie, Université de Genève, Chemin Pegasi 51, CH-1290 Versoix, Switzerland}
\affiliation{Gravitational Wave Science Center (GWSC), Université de Genève, CH1211 Geneva, Switzerland}

\author[0000-0003-1474-1523]{Tassos\,Fragos}
\affiliation{Département d’Astronomie, Université de Genève, Chemin Pegasi 51, CH-1290 Versoix, Switzerland}
\affiliation{Gravitational Wave Science Center (GWSC), Université de Genève, CH1211 Geneva, Switzerland}

\author[0000-0001-6692-6410]{Seth\,Gossage}
\affiliation{Center for Interdisciplinary Exploration and Research in Astrophysics (CIERA), Northwestern University, 1800 Sherman Ave, Evanston, IL 60201, USA}
\affiliation{NSF-Simons AI Institute for the Sky (SkAI),172 E. Chestnut St., Chicago, IL 60611, USA}

\author[0000-0003-3684-964X]{Konstantinos\,Kovlakas}
\affiliation{Institute of Space Sciences (ICE, CSIC), Campus UAB, Carrer de Magrans, 08193 Barcelona, Spain}
\affiliation{Institut d'Estudis Espacials de Catalunya (IEEC),  Edifici RDIT, Campus UPC, 08860 Castelldefels (Barcelona), Spain}

\author[0000-0001-9331-0400]{Matthias\,U.\,Kruckow}
\affiliation{Département d’Astronomie, Université de Genève, Chemin Pegasi 51, CH-1290 Versoix, Switzerland}
\affiliation{Gravitational Wave Science Center (GWSC), Université de Genève, CH1211 Geneva, Switzerland}

\author[0000-0003-4474-6528]{Kyle\,A.\,Rocha}
\affiliation{Department of Physics and Astronomy, Northwestern University, 2145 Sheridan Road, Evanston, IL 60208, USA}
\affiliation{Center for Interdisciplinary Exploration and Research in Astrophysics (CIERA), Northwestern University, 1800 Sherman Ave, Evanston, IL 60201, USA}
\affiliation{NSF-Simons AI Institute for the Sky (SkAI),172 E. Chestnut St., Chicago, IL 60611, USA}

\author[0000-0003-1749-6295]{Philipp\,M.\,Srivastava}
\affiliation{Electrical and Computer Engineering, Northwestern University, 2145 Sheridan Road, Evanston, IL 60208, USA}
\affiliation{Center for Interdisciplinary Exploration and Research in Astrophysics (CIERA), Northwestern University, 1800 Sherman Ave, Evanston, IL 60201, USA}
\affiliation{NSF-Simons AI Institute for the Sky (SkAI),172 E. Chestnut St., Chicago, IL 60611, USA}

\author[0000-0001-9037-6180]{Meng\,Sun}
\affiliation{Center for Interdisciplinary Exploration and Research in Astrophysics (CIERA), Northwestern University, 1800 Sherman Ave, Evanston, IL 60201, USA}

\author[0000-0003-0420-2067]{Elizabeth\,Teng}
\affiliation{Department of Physics and Astronomy, Northwestern University, 2145 Sheridan Road, Evanston, IL 60208, USA}
\affiliation{Center for Interdisciplinary Exploration and Research in Astrophysics (CIERA), Northwestern University, 1800 Sherman Ave, Evanston, IL 60201, USA}
\affiliation{NSF-Simons AI Institute for the Sky (SkAI),172 E. Chestnut St., Chicago, IL 60611, USA}

\author[0000-0002-0031-3029]{Zepei\,Xing}
\affiliation{Département d’Astronomie, Université de Genève, Chemin Pegasi 51, CH-1290 Versoix, Switzerland}
\affiliation{Gravitational Wave Science Center (GWSC), Université de Genève, CH1211 Geneva, Switzerland}

\author[0000-0002-7464-498X]{Emmanouil\,Zapartas}
\affiliation{Institute of Astrophysics, Foundation for Research and Technology-Hellas, GR-71110 Heraklion, Greece}

\begin{abstract}
Binaries harboring a \ac{MSP} and a \ac{BH} are a key observing target for current and upcoming pulsar surveys. 
 We model the formation and evolution of such binaries in isolation at solar metallicity using the next-generation binary population synthesis code \texttt{POSYDON}.
 We examine \ac{NS}--\ac{BH} binaries where the NS forms first (labeled NSBH), as the \ac{NS} must be able to spin--up to \ac{MSP} rotation periods before the \ac{BH} forms in these systems.
 We find that NSBHs are very rare and have a birth rate $<$ 1 Myr$^{-1}$ for a Milky Way-like galaxy in our typical models.
 The NSBH formation rate is 2--3 orders of magnitude smaller than that for \ac{NS}--\acp{BH} where the \ac{BH} forms first (labeled BHNS).  
 These rates are also sensitive to model assumptions about the \ac{SN} remnant masses, natal kicks, and \ac{CE} efficiency.
 We find that 100\% of NSBHs undergo a mass ratio reversal before the first \ac{SN} and up to 64\% of NSBHs undergo a double common envelope phase after the mass ratio reversal occurs. 
 Most importantly, no NSBH binaries in our populations undergo a mass transfer phase, either stable or unstable, after the first \ac{SN}.
 This implies that there is no possibility of pulsar spin-up via accretion, and thus \ac{MSP}--\ac{BH} binaries cannot form.
 Thus, dynamical environments and processes may provide the only formation channels for such \ac{MSP}--\ac{BH} binaries. 
 
\end{abstract}

\section{Introduction}\label{sec:intro}

Pulsars are highly magnetized, rapidly rotating \acp{NS} most commonly detectable as radio sources. 
Some pulsars get ``recycled" during their lifetime via accretion, which causes them to spin--up to shorter rotation periods.
If they reach periods $\lesssim$~30~ms, this places them in the distinct category of \acp{MSP}~\citep[see][for reviews]{phinney_binary_1994, lorimer_binary_2008, dantona_origin_2020}.

Detecting a Galactic pulsar--\ac{BH} binary is an observation goal for current and upcoming pulsar surveys, such as the ongoing MeerKAT~\citep{booth_meerkat_2009}, FAST~\citep{nan_five-hundred-meter_2011}, and CHIME~\citep{stairs_observing_2019, amiri_chime_2021} surveys, as well as surveys with the upcoming Square Kilometer Array~\citep{braun_anticipated_2019}.
Observing a pulsar--\ac{BH} binary would enable unprecedented tests of general relativity due to the extreme curvature of spacetime around the system~\citep[e.g.,][]{kramer_strong-field_2004, seymour_testing_2018}.
A \ac{MSP}--\ac{BH} detection is of particular interest because the short \ac{MSP} pulse periods provide the timing precision necessary to measure the Shapiro delay (the signal delay resulting from light passing a massive object), which can provide additional constraints on the \ac{NS} mass and equation of state~\citep{shapiro_fourth_1964, fonseca_fundamental_msp_2019}. 
In addition, \acp{MSP} can exist as detectable sources for over ten times longer than regular pulsars and have wider beams due to their short pulse periods, both of which increase their overall detection probability \citep[e.g.,][]{lorimer_binary_2008}.

A recent observation made with the MeerKAT telescope has identified a \ac{MSP} in a binary with a \ac{CO} in the lower mass gap between the heaviest \acp{NS} and the lightest \acp{BH}~\citep{barr_pulsar_2024}.
Located in the \ac{GC} NGC 1851, this system is thought to have formed dynamically~\citep{barr_pulsar_2024}; the \ac{MSP} was spun--up by a low-mass companion that was later replaced by a high-mass companion in an exchange interaction.
The lack of radio detections of the high-mass companion makes it impossible to distinguish between a high-mass \ac{NS} and low-mass \ac{BH}~\citep{barr_pulsar_2024}.

Many studies have predicted how \ac{MSP}--\ac{BH} binaries, as well as \ac{NS}--\ac{BH} binaries more broadly, could form dynamically in \acp{GC}~\citep[e.g.,][]{clausen_dynamically_2014, ye_rate_2019, ye_millisecond_2019, arca_sedda_dissecting_2020, fragione_demographics_2020}.
While it is known that \acp{GC} are very efficient at forming \acp{MSP} compared to the Galactic field due to their high stellar densities~\citep[e.g.,][]{manchester_australia_2005, ransom_pulsars_2008, bahramian_slar_2013}, \acp{GC} are relatively inefficient at forming \ac{NS}--\ac{BH} binaries~\citep{clausen_black_2013, ye_rate_2019, arca_sedda_dissecting_2020, hoang_neutron_2020}, and the number of \acp{MSP} may even be anti-correlated with the number of \acp{BH} in a given cluster~\citep{ye_millisecond_2019}.
In this study, we focus on the formation of isolated \ac{MSP}--\ac{BH} binaries in the Galactic field and thus do not examine the possible dynamical formation channels of these systems.

Many \ac{BPS} studies have previously examined isolated \ac{NS}--\ac{BH} formation with a focus on their properties as \ac{GW} sources~\citep{kruckow_progenitors_2018, mapelli_cosmic_2018, shao_black_2018, broekgaarden_impact_2021, chattopadhyay_modelling_2021, roman-garza_role_2021, drozda_black_2022, xing_zams_2023, xing_mass-gap_2024}.
Growing interest in this area of research is related to the \ac{GW} detections consistent with being the products of \ac{NS}--\ac{BH} coalescences that have been recently reported by the LIGO--Virgo--KAGRA (LVK) collaboration.
Their connection to multi-messenger astronomy as potential gamma-ray burst and kilonovae sources is also of particular interest \citep[see e.g.,][for reviews]{meszaros_multi-messenger_2019, branchesi_multi-messenger_2023}. 

The events GW200115 and GW230529 were both reported with high confidence and with estimated component masses consistent with \ac{NS}--\ac{BH} binaries ~\citep{abbott_observation_2021,abac_observation_2024}. However, the most massive (primary) component for GW230529 was estimated between $2.5$--$4.5$~$M_{\odot}$, and one cannot rule out that it was a massive \ac{NS}. GW190814 was also reported with high confidence, with primary and secondary masses of $23.3^{+1.4}_{-1.4}$ and $ 2.6^{+.1}_{-.1}$~$M_{\odot}$, respectively ~\citep{abbott_gw190814_2020, abbott_gwtc-21_2024}. Given the high secondary mass and the absence of additional tidal measurements or electromagnetic counterparts, it remains unclear whether GW190814 originated from a \ac{NS}--\ac{BH} or \ac{BH}--\ac{BH} coalescence. GW200210 was detected with a similar secondary mass of $2.83^{+.47}_{-.42}$~$M_{\odot}$ but a lower probability of astrophysical origin of 0.54 ~\citep{abbott_gwtc-3_2023}. GW191219 and GW190917 were reported with component masses consistent with \ac{NS}--\ac{BH} binaries, but with uncertain probabilities of astrophysical origin due to systematics ~\citep{abbott_gwtc-3_2023, abbott_gwtc-21_2024}. Finally, GW190426\_152155 and GW200105 were reported as marginal candidates with masses consistent with \ac{NS}--\ac{BH} binaries but low probability ($<0.5$) of astrophysical origin ~\citep{abbott_observation_2021,abbott_gwtc-3_2023, abbott_gwtc-21_2024}.

In this paper, we use the \texttt{POSYDON} population synthesis code~\citep{fragos_posydon_2023, andrews_posydon_2024} to examine the possible formation channels of isolated Galactic \ac{NS}--\ac{BH} binaries that host recycled \acp{NS}, regardless of whether they become potential \ac{GW} sources or not.
As previously mentioned, the \acp{NS} in these binaries are expected to form first so that they may accrete from their companions and be spun--up before the \ac{BH} forms.
Past \ac{BPS} studies have examined such binaries in detail, including \citet{kruckow_progenitors_2018} and \citet{chattopadhyay_modelling_2021}, but come to opposing conclusions regarding their presence in Milky Way-like galaxies.
We discuss our detailed results in conversation with these findings in Section~\ref{sec:study_comparison}.

In Section~\ref{sec:methods}, we discuss our \ac{BPS} models and the \texttt{POSYDON} code.
In Section~\ref{sec:results}, we discuss the properties of \ac{NS}--\ac{BH} binaries in our \ac{BPS} populations, including their birth rates, formation channels, and \ac{DCO} properties.
In Section~\ref{sec:discussion}, we place our findings in the context of previous studies as well as current and future \ac{MSP}--\ac{BH} detections, in addition to discussing our model uncertainties.
In Section~\ref{sec:conclusion}, we summarize our conclusions.

\section{Methods}\label{sec:methods}
We generate populations of $10^7$ binaries at solar metallicity ($Z$ = 0.0142) with Version 2 of the \ac{BPS} code \texttt{POSYDON}~\citep{andrews_posydon_2024}.
\texttt{POSYDON} is unique from other \ac{BPS} codes in that it uses detailed grids of stellar and binary models generated using \texttt{MESA} \citep{paxton_modules_2011, paxton_modules_2013, paxton_modules_2015, paxton_modules_2019} to self-consistently evolve the binary components' stellar structure with the binary evolution. 
For a detailed description of the code, see \citet{fragos_posydon_2023} and \citet{andrews_posydon_2024}.

\subsection{Binary Population Models}
We use \texttt{POSYDON}'s initial-final interpolation scheme~\citep{fragos_posydon_2023, andrews_posydon_2024} to acquire binary properties from the grids of pre-computed models in all of our populations.
This scheme first classifies binaries being evolved with a given \texttt{MESA} grid into a \ac{MT} history class for that grid based on the binary component masses and orbital period.
Then, the physical properties of the binaries are interpolated from the grid values within each \ac{MT} class~\citep{fragos_posydon_2023, andrews_posydon_2024}.

The BPS initial conditions are as follows: We draw primary masses
using the \ac{IMF} from \citet{kroupa_variation_2001} and initial binary orbital periods using the prescription detailed in \citet{sana_binary_2012}. 
Primary masses are in the range 6.2--120~$M_{\odot}$ and secondary masses are in the range 0.35--120~$M_{\odot}$ to align with the mass ranges of \texttt{POSYDON}'s  pre-computed binary grids. 
We sample secondary masses assuming an initially flat mass ratio distribution. 
We apply a burst star formation history, which assumes all stars are formed at the same time, and evolve our binaries for 13.8~Gyr. We obtain astrophysical results by renormalizing to a constant star formation rate. 

We make several additional modeling assumptions, detailed below, including those that affect \texttt{POSYDON}'s ``on-the-fly" calculations for stages of evolution not handled by the pre-calculated \texttt{MESA} grids.

\subsection{Mass Transfer}\label{sec:mass_transfer}

All \ac{MT} in \texttt{POSYDON} is handled within the pre-computed \texttt{MESA} grids of binary models, and thus \ac{MT} settings cannot be changed without running new grids.
We briefly describe how different \ac{MT} cases are treated in the standard \texttt{POSYDON} grids used for this study.
For full details, refer to \citet{fragos_posydon_2023} and \citet{andrews_posydon_2024}.

Mass-loss due to \ac{RLO} of \ac{MS} stars is handled using the \texttt{contact} scheme within \texttt{MESA}.
This prescription is suitable for modeling the structure of \ac{MS} binary evolution, including cases in which one or both stars overfill their Roche lobes. 
When stars evolve off the \ac{MS}, modeling is switched to the \texttt{Kolb} scheme~\citep{kolb_comparative_1990}, which more accurately handles stars with expanded envelopes.

\texttt{POSYDON} Version 2 also includes treatment for reverse \ac{MT}, which occurs when a \ac{MT} phase from the primary to the secondary takes place and afterwards the secondary initiates ``reverse" \ac{MT} back onto the primary.
This happens frequently in regions of parameter space where the \ac{ZAMS} mass ratio of the binary is close to one.
Reverse \ac{MT} is not traditionally supported by the \texttt{Kolb} scheme, and thus modifications were made to the \texttt{MESA} code used to run \texttt{POSYDON}'s pre-calculated grids that enable the switching of donors when the \ac{MT} rate from the secondary exceeds that of the primary~\citep{andrews_posydon_2024}. 

Mass accretion is handled differently depending on the accretor type.
For binaries with a non-degenerate accretor, all the mass lost by the donor through \ac{RLO} is initially accepted by the accretor.
This spins-up the accreting star as long as it has not reached a critical rotation rate.
In the case of critically-rotating stars, accretion is restricted and mass will begin to be ejected by the binary through isotropic stellar winds.
The wind mass-loss rate of the accretor is calculated with a rotationally-enhanced wind model that boosts stellar winds to ensure the stellar rotation rate always remains below its critical threshold~\citep{fragos_posydon_2023}.
The overall effect of these assumptions is that the mass accretion may be highly inefficient depending on the accretor's angular momentum and stellar structure, even when enforcing that all mass lost from the donor is initially transferred to the accretor during \ac{RLO}.
While boosted winds are commonly used to model these systems ~\citep[e.g.,][]{paxton_modules_2013}, star-disk interactions also play a crucial role in the accretor spin-up.
Improved understanding of star-disk interactions could change the spin-up and accretion assumptions for rapidly-rotating stars~\citep{colpi_analytical_1991, paczynski_polytropic_1991, popham_does_1991}.

For binaries with degenerate accretors, \ac{MT} proceeds similarly except that it is Eddington-limited; sub-Eddington \ac{MT} rates are conservative, and super-Eddington rates cause excess matter to be lost via an isotropic wind from the vicinity of the accretor.
For full details on how the Eddington-limited accretion rate is calculated, refer to ~\citet{fragos_posydon_2023}

\subsection{Common-Envelope Evolution}\label{sec:CE}
If a binary experiences dynamically unstable \ac{MT}, it can undergo \ac{CE} evolution~\citep{paczynski_common_1976, ivanova_common_2020}. 
\texttt{POSYDON} determines the onset of an unstable \ac{MT} phase using various criteria in its pre-computed grids of binary models.
First, if the \ac{MT} rate of the binary exceeds 0.1 $M_{\odot}$ $\mathrm{yr}^{-1}$, the binary is assumed to enter an unstable \ac{RLO} phase.
The binary will also begin unstable \ac{MT} if the stellar radius of the expanding star extends beyond the gravitational equipotential surface, or second Lagrangian point L$_2$. 
For \ac{CO} accretors, a \ac{CE} phase is assumed to begin if the accretor's photon-trapping radius reaches its Roche-lobe radius. 
Lastly, if both stars in the binary fill their Roche-lobes while at least one of the stars is in a post-\ac{MS} phase of evolution, the binary is assumed to initiate unstable \ac{MT}.
For full details on these assumptions, refer to~\citet{fragos_posydon_2023}.

We treat \ac{CE} as an instantaneous process using the standard $\alpha$ -- $\lambda$ formalism \citep{webbink_double_1984, livio_common_1988}, where $\alpha$ is the \ac{CE} efficiency parameter and $\lambda$ is the envelope binding energy factor.  
We set $\alpha = 1$ for the efficiency parameter in our default models, but also examine model variations with $\alpha = 0.5$ and $\alpha = 0.1$ in Section~\ref{sec:alpha}.
For the binding energy factor, \texttt{POSYDON} is able to self-consistently calculate $\lambda$ at the onset of \ac{CE} using the stellar profiles from the pre-calculated binary grids.
To define the core-envelope boundary, we assume a \ac{H}  abundance fraction and a \ac{He} abundance fraction of 0.1.
If a \ac{CE} ejection is successful, we assume that that core mass and radius of the star are the same as before the CE began, rather than being set by the \ac{H} and \ac{He} abundance fractions. 
We assume that Hertzsprung-Gap donor stars are able to undergo and survive a \ac{CE} phase, which is commonly referred to as the ``optimistic" \ac{CE} scenario in the literature~\citep[e.g.,][]{vigna-gomez_formation_2018}.

\texttt{POSYDON} also includes treatment for double \ac{CE} evolution.
This occurs when both stars in the binary have overfilled their Roche-lobes and both have a giant-like structure with a distinct core-envelope separation.
In addition, unstable \ac{MT} must have been initiated by at least one star in the binary.
In this case, $\lambda$ is calculated separately for both stars ($\lambda_1$, $\lambda_2$) and the binding energy for each is summed to get the total binding energy:
\begin{equation}
    E_{\mathrm{bind}} = -G\left[\frac{m_1(m_1 - m_{\mathrm{1, c}})}{\lambda_1 R_1} + \frac{m_2(m_2 - m_{\mathrm{2, c}})}{\lambda_2 R_2}\right]
\end{equation}
where $m_i$, $m_{i,\mathrm{c}}$, and $R_i$ are the total stellar mass, core mass, and stellar radius of the $i$-th component of the binary, respectively.
The orbital energy and separation of the binary after the double \ac{CE} are then calculated using $E_{\mathrm{bind}}$ like in the single \ac{CE} case~\citep{fragos_posydon_2023}.

Though the modeling of double \ac{CE} is uncertain, there is observational evidence indicating that this process does occur in nature~\citep{ahmad_discovery_2004, justham_formation_2011, sener_spectroscopic_2014}.
The conditions for the onset of a double \ac{CE} are thought to differ from those for a single \ac{CE} because \textit{both} stars in a double \ac{CE} binary have well-developed cores with extended envelopes (in the traditional single \ac{CE} treatment, one star is expected to have this structure while the other object is assumed to be dense).
This implies that the stars have comparable masses, in which case \ac{MT} is not expected to become dynamically unstable as a direct consequence of \ac{RLO} or tidal instability~\citep{ivanova_common_2020}.
There also may be additional methods of initiating unstable \ac{MT} particular to these binaries that would require expensive 3D hydrodynamical simulations to understand, which has not yet been done~\citep{ivanova_common_2020}.
Thus, it is uncertain how many binaries whose stars meet our double \ac{CE} requirements actually initiate unstable \ac{MT} and reach orbital instability, and of these binaries, which will undergo a successful double \ac{CE} phase and which will merge.

\subsection{Core Collapse}\label{sec:supernovae}
The model used to calculate the remnant masses of \ac{CO}s formed from the core collapse of massive stars significantly impacts the \ac{NS} and \ac{BH} mass distributions in binary populations.
We model binary populations using the ``delayed" prescription from~\citet{fryer_compact_2012} as well as the ``N20" engine prescription from~\citet{sukhbold_core-collapse_2016}, as these are two prescriptions in \texttt{POSYDON} that allow for the formation of \acp{BH} in the low-end mass gap (3--5~$M_{\odot}$). 
This is important to consider given the recent observational evidence of \acp{BH} existing in this gap~\citep{abac_observation_2024}.
The \ac{F12d} model uses the \ac{C--O} core mass of the star at the start of the \ac{CC} as well as the amount of fallback material onto the proto-\ac{CO} after the explosion to determine the final baryonic mass of the \ac{CO}.
This prescription also uses a maximum \ac{NS} mass limit to determine the remnant type (\ac{NS} or \ac{BH}), which we set to 2.5~$M_{\odot}$.
The \ac{S16} model uses the \ac{He} core mass of the star prior to its collapse to determine both the remnant mass and fallback onto the \ac{CO} as well as its remnant type.

Stars with \ac{ZAMS} masses in the range $\sim$7--10~$M_{\odot}$ may undergo an \ac{ECSN} when they have degenerate ONeMg cores, producing less-energetic explosions and hence smaller \ac{SN} kicks~\citep[e.g.,][]{nomoto_evolution_1984, podsiadlowski_effects_2004, hiramatsu_electron-capture_2021}.
For \acp{ECSN}, we model binary populations using the prescription from~\citet{podsiadlowski_effects_2004}.
With this formalism, pre-\ac{CC} stars with core \ac{He} masses between 1.4--2.5~$M_{\odot}$ will undergo an \ac{ECSN}.

Stars that have lost their surface layers during binary interactions may undergo an \ac{USSN} and will receive smaller natal kicks than in a \ac{CCSN}~\citep[e.g.,][]{tauris_ultra-stripped_2015}.
\texttt{POSYDON} currently does not include treatment for \acp{USSN}.
However, we do not expect this to affect our main results, as past studies have shown that \acp{USSN} primarily occur in \ac{NS}--\ac{BH} binaries where the \ac{BH} forms first~\citep{chattopadhyay_modelling_2021}.

To model the kicks of newly-born \ac{CO}s formed from \acp{CCSN}, we sample kicks from a Maxwellian distribution with a velocity dispersion of 265~km~s~$^{-1}$~\citep{hobbs_statistical_2005}.
For the kicks of \acp{NS} formed from \acp{ECSN}, we use a velocity dispersion of 20~km~s~$^{-1}$~\citep{giacobbo_impact_2019}.
We also investigate one model variation with reduced \acp{CCSN} kicks,  designated as ``low $\sigma_{\mathrm{CCSN}}$", that uses the velocity dispersion estimates from~\citet{mandel_simple_2020}.
This study applies the results from three-dimensional supernova simulations and semi-analytical parameterized models to derive stochastic prescriptions for the natal kicks of \acp{NS} and \acp{BH}. 
They estimate that the scatter for \ac{NS} kicks is $\sim$120 km s$^{-1}$ and the scatter for \ac{BH} kicks is $\sim$60 km s$^{-1}$.
They acknowledge that these estimates under-predict kick velocities for the least-massive \ac{C--O} cores ($\lesssim$ 4~$M_{\odot}$) compared to detailed semi-analytic predictions due to uncertainties in stellar evolution and explosion models~\citep{mandel_simple_2020}.
For this reason, we consider their kick scatter estimates to be a lower-limit for natal kicks, particularly for \acp{NS} formed from low-mass \ac{C--O} cores.

As in the default \texttt{POSYDON} treatment, we normalize all \ac{BH} kicks by multiplying by 1.4~$M_{\odot}$ and dividing by the remnant mass in order to approximately rescale the \ac{NS} kick distribution to one suited for heavier \acp{CO}~\citep{fragos_posydon_2023}.

\section{Results}\label{sec:results}

We examine the subpopulations of \ac{NS}--\ac{BH} binaries in which the \ac{NS} forms first (NSBHs) in contrast to those where the \ac{BH} forms first (BHNSs). 
We focus on the properties and evolution of NSBH binaries, as we are interested in their relation to \ac{MSP}--\ac{BH} systems in which it is necessary for the \ac{NS} to form first so that it has the chance to be spun-up to millisecond rotation periods before \ac{BH} formation.

\subsection{Frequency of NS--BH Binaries}\label{sec:frequencies}

\begin{deluxetable}{c c c c}
\tablecaption{Birth rates (Myr$^{-1}$) of NSBH, BHNS, and combined NS--BH binaries in a Milky Way-like galaxy for all of our population models. 
}
\label{table:frequencies}
\tablehead{
\colhead{Model} &  \colhead{\hspace{.75cm}NSBH}  &  \colhead{\hspace{.75cm}BHNS} &
\colhead{\hspace{.75cm}NS--BH}\hspace{.5cm}
}
\startdata
F12d & \hspace{.75cm}0.05 & \hspace{.75cm}36.35  & \hspace{.45cm}36.40\\
S16 & \hspace{.75cm}0.63 & \hspace{.75cm}63.90 & \hspace{.45cm}64.53\\
F12d, low $\sigma_{\mathrm{CCSN}}$ & \hspace{.75cm}0.30 & \hspace{.75cm}125.31 & \hspace{.45cm}125.61\\
S16, low $\sigma_{\mathrm{CCSN}}$ & \hspace{.75cm}1.61 & \hspace{.75cm}174.63 & \hspace{.45cm}176.24\\
S16, $\alpha$ = 0.5 & \hspace{.75cm}0.24 & \hspace{.75cm}55.39  & \hspace{.45cm}55.63\\
S16, $\alpha$ = 0.1 & \hspace{.75cm}0.19 & \hspace{.75cm}50.95  & \hspace{.45cm}51.14\\
\enddata
\end{deluxetable}

\begin{figure}
\includegraphics[width=0.5\textwidth]{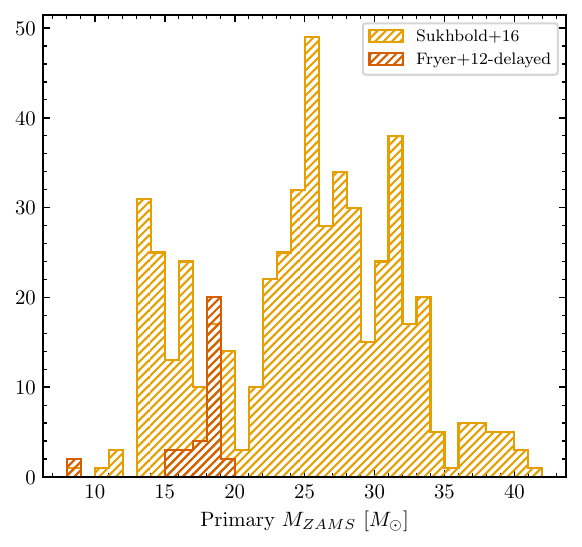}
\caption{
The primary \ac{ZAMS} masses of all NSBH progenitors in our  \ac{S16} and \ac{F12d} populations.
The differing mass ranges between prescriptions is a result of how each computes remnant masses and assigns \ac{CO} types (Section~\ref{sec:frequencies}).
}
\label{fig:ZAMS_masses}
\end{figure}

\begin{figure}
\includegraphics[width=0.5\textwidth]{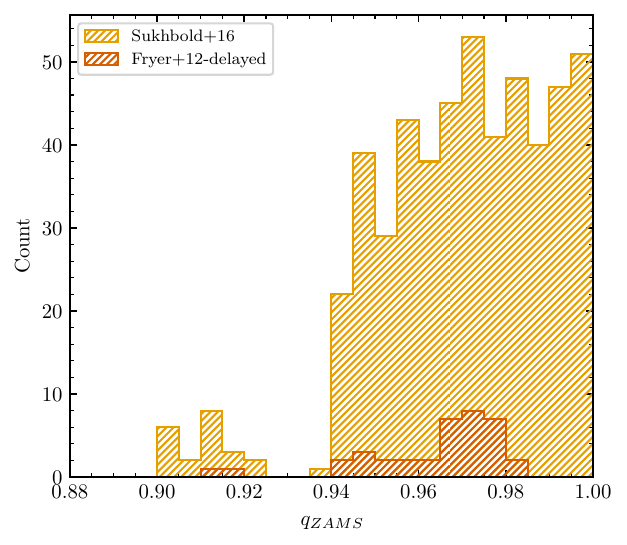}
\caption{The mass ratio distributions at \ac{ZAMS} ($q_{ZAMS}$) for all NSBH progenitors. 
In both the \ac{F12d} and  \ac{S16} populations, the \ac{ZAMS} mass ratios are close to unity.}
\label{fig:mass_ratios}
\end{figure}

We calculate the birth rates of NSBHs and BHNSs in a Milky Way-like galaxy to understand their relative frequencies in our populations (Table~\ref{table:frequencies}).
We calculate birth rates by rescaling the number of binaries in a given subpopulation with the average star formation rate of the Milky Way, which we take to be SFR$_{MW}$ = 1.65~$M_{\odot}/\mathrm{yr}$~\citep{licquia_improved_2015}.
We normalize our simulations by: multiplying the number of systems by SFR$_{MW}$ and dividing by the total underlying mass of the \texttt{POSYDON} population, which is the stellar mass of the population at \ac{ZAMS} after integrating the associated population \ac{IMF} and period distribution fully.
Our \ac{NS}--\ac{BH} birth rates are order-of-magnitude comparable with those calculated by \citet{chattopadhyay_modelling_2021}.

In most of our \texttt{POSYDON} models, we find that the birth rate of bound NSBH binaries, including systems that do and do not merge in a Hubble time, is $<$~1~Myr$^{-1}$.
In all of our populations, the birth rate of NSBH binaries is 2--3 orders of magnitude smaller than that for BHNS binaries.
This is to be expected, as \acp{BH} tend to form from heavier stellar primaries that evolve more quickly than their lighter secondary counterparts.
The high prevalence of BHNSs compared to NSBHs is in agreement with previous \ac{BPS} studies that examine \ac{NS}--\ac{BH} populations~\citep[e.g.,][]{kruckow_progenitors_2018, chattopadhyay_modelling_2021, xing_zams_2023}.

When comparing the birth rates of NSBHs in the  \ac{F12d} vs. the  \ac{S16} populations, we find that there are about 10 times more NSBHs in the  \ac{S16} population, and the total \ac{NS}--\ac{BH} rate increases by $\lesssim 2$.
This is a result of differences in how remnant masses are determined in the \ac{F12d} and  \ac{S16} prescriptions.

The \ac{S16} \ac{SN} mechanism, exploded with the N20 engine in our populations, produces \ac{CO} remnant types in a relatively stochastic manner.
In \citet{sukhbold_core-collapse_2016}, they find that N20 models with \ac{ZAMS} stellar masses up to 120~$M_{\odot}$ can successfully explode and form \ac{NS} remnants, while models with \ac{ZAMS} masses as low as $\sim$15~$M_{\odot}$ can implode to form \ac{BH} remnants (not taking into account binary interactions that may change the stellar mass before collapse).
The primary \ac{ZAMS} mass parameter space for our NSBH binaries in the \ac{S16} population falls between 10--40~$M_{\odot}$ (Figure~\ref{fig:ZAMS_masses}) and the secondary masses fall in a similar range, as all NSBHs have \ac{ZAMS} mass ratios close to one (Figure~\ref{fig:mass_ratios}, see \ref{sec:form_channels} and \ref{sec:study_comparison} for discussion).
The remnant types produced for stars with primary \ac{ZAMS} masses between 15--40~$M_{\odot}$ switches frequently between \acp{BH} and \ac{NS} in the \ac{S16} treatment~\citep{sukhbold_core-collapse_2016}.
Thus, stars with higher initial masses can end-up producing \ac{NS} remnants, and those with lower initial masses can produce \ac{BH} remnants, which constructs a wider parameter space for possible NSBH formation.

The \ac{F12d} mechanism functions differently, as it determines the remnant \ac{CO} type using the maximum \ac{NS} mass limit, which we set to 2.5~$M_{\odot}$.
The remnant mass itself is determined using the \ac{C--O} core mass of the progenitor star at the time of explosion, which is related to the mass of the star at \ac{ZAMS}~\citep{fryer_compact_2012}.
Because all NSBH progenitors have \ac{ZAMS} mass ratios close to one (Section~\ref{sec:form_channels}, Figure~\ref{fig:mass_ratios}), this implies that, when evolved with the \ac{F12d} prescription, they are only likely to have masses that fall in the necessary range for becoming high-mass \acp{NS} or low-mass \acp{BH}.
This range corresponds to \ac{ZAMS} masses between $\sim$15--20~$M_{\odot}$ according to the results in~\citet{fryer_compact_2012}.
Indeed, we find that 86\% of the NSBHs in the \ac{F12d} population have primary \ac{ZAMS} masses in this range (Figure~\ref{fig:ZAMS_masses}).
Because the parameter space for possible NSBH formation is narrower in the \ac{F12d} treatment, much fewer NSBHs will form in a given population.

\subsection{NSBH Formation Channels}\label{sec:form_channels}

\begin{figure}
\includegraphics[width=0.5\textwidth]{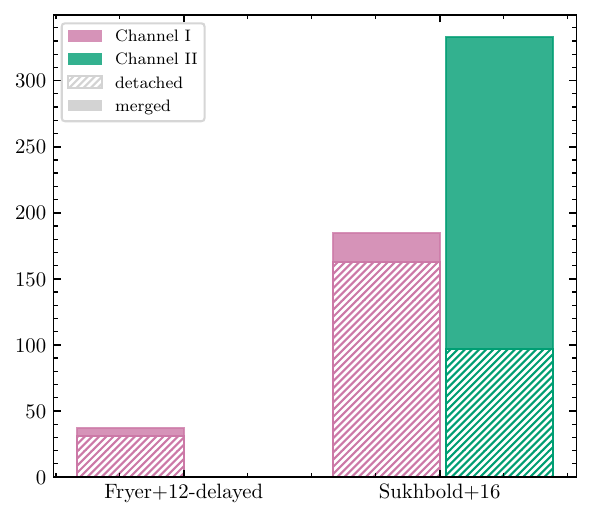}
\caption{
The number of NSBH binaries in the \ac{F12d} and  \ac{S16} populations that form through Channel~I (pink) vs. Channel~II (green) (Sections~\ref{sec:channel_I}~\&~\ref{sec:channel_II}). 
We show the subset of binaries in each channel that merge in a Hubble time (solid) and those that do not merge, or remain detached (hatched).
}
\label{fig:form_channels}
\end{figure}

As previously mentioned, all NSBH progenitors in our populations have \ac{ZAMS} mass ratios ($q = M_2/M_1$) close to unity ($\gtrsim$~0.9) as shown in Figure~\ref{fig:mass_ratios}.
We find that binaries with \ac{ZAMS} mass ratios down to $\sim$~0.4 can have primaries that form \acp{NS} and secondaries that form \acp{BH}, but they get disrupted at the first \ac{SN} and thus do not contribute to our NSBH populations.

Having \ac{ZAMS} mass ratios close to one also makes it easier for these systems to undergo a mass ratio reversal (in which the secondary object becomes heavier than the primary) before the first \ac{SN}, which increases the possibility of \ac{NS} formation occurring before \ac{BH} formation depending on the \ac{SN} prescription and \ac{ZAMS} mass values.
We find that 100\% of NSBH progenitors in both our \ac{S16} and \ac{F12d} populations have a heavier secondary star than primary star before the first \ac{SN} due to \ac{MT} during their \ac{HMS--HMS} evolution (Sections~\ref{sec:hms_hms} and \ref{sec:study_comparison}).

We find that NSBH binaries form via two different channels after the \ac{HMS--HMS} phase. 
The first channel, designated as Channel I, has fully detached evolution of the binary components after the \ac{HMS--HMS} phase. 
In the second channel, designated as Channel II, the binaries undergo a double \ac{CE} phase immediately after the \ac{HMS--HMS} phase and before the first \ac{SN}.
The breakdown of how many binaries form through each channel in each population, including binaries that do and do not merge in a Hubble time, is shown in Figure~\ref{fig:form_channels}.
We discuss these two different channels in detail in Sections~\ref{sec:channel_I}~\&~\ref{sec:channel_II}.

\subsubsection{Mass Transfer Before First \ac{SN}}\label{sec:hms_hms}

\begin{figure*}
\includegraphics[width=\textwidth]{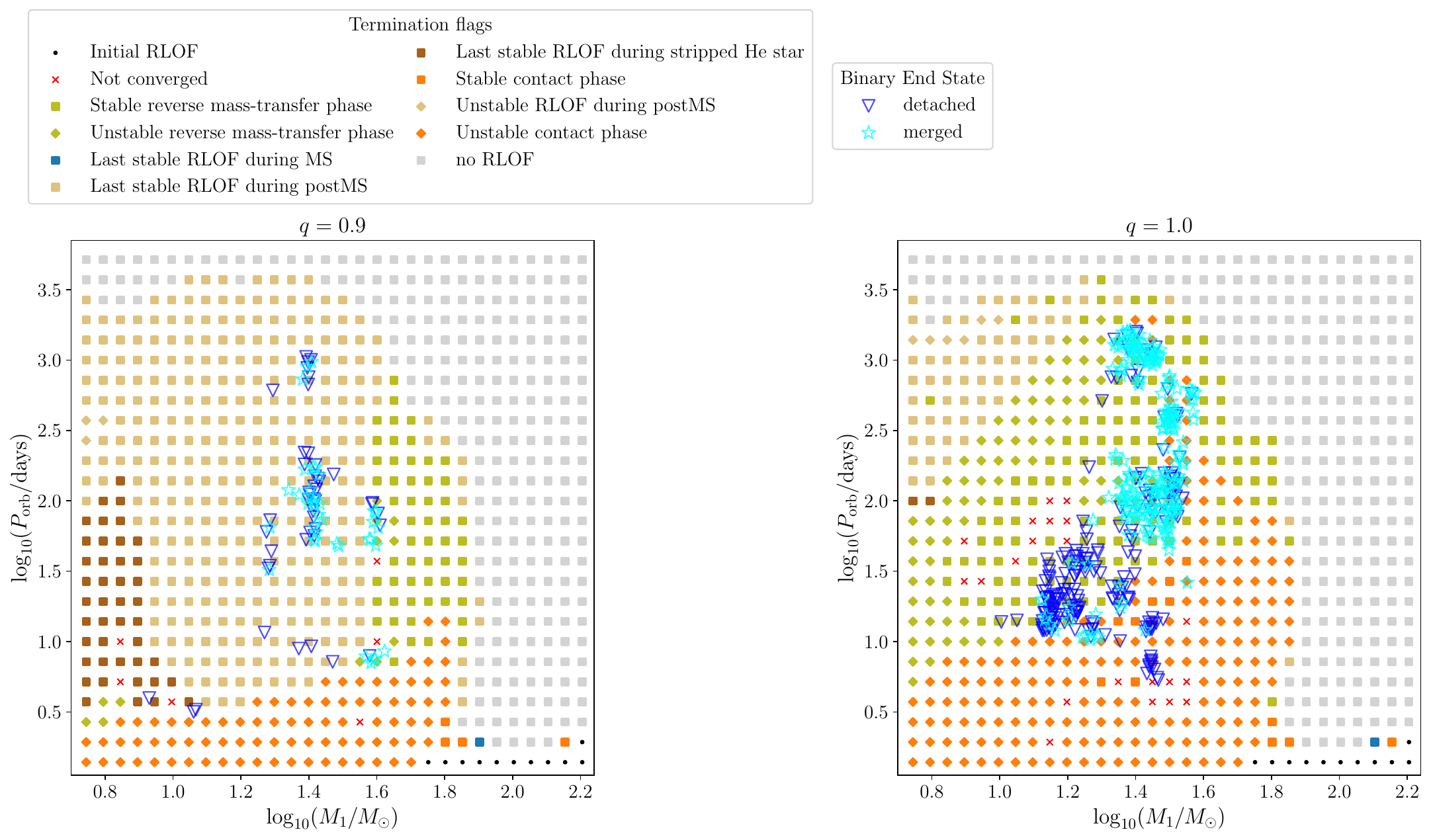}
\caption{The \ac{HMS--HMS} grid slices from \texttt{POSYDON}'s library of binary star models for \ac{ZAMS} mass ratios $q=0.9$ and $q=1.0$.
The different grid symbols summarize the evolution of each of the pre-computed \texttt{MESA} models. 
The \ac{ZAMS} binary orbital periods vs. primary masses for all NSBH progenitors in the  \ac{S16} \ac{BPS} population are plotted over the grids, where the cyan star and blue triangle markers distinguish between binaries that do and do not merge in a Hubble time, respectively. 
Binaries with initial mass ratios $\leq$ 0.95 are potted on the $q=0.9$ grid slice, while those with mass ratios $>$ 0.95 are plotted on the $q=1.0$ grid slice.}
\label{fig:S16_grids}
\end{figure*}

We examine the \ac{MT} history before a possible double \ac{CE} is triggered and before the first \ac{SN} by looking at the \ac{HMS--HMS} \texttt{MESA} grid slices from \texttt{POSYDON}'s library of binary star models.
The \ac{HMS--HMS} grids summarize the evolution of binaries with two stars at \ac{ZAMS} until they reach a designated stopping condition in \texttt{MESA}~\citep{fragos_posydon_2023, andrews_posydon_2024}.
Each grid is computed for a range of initial primary masses ($M_1$), orbital periods ($P_{\mathrm{orb}}$), and mass ratios ($q$).
An example of these grid slices overlaid with the \ac{ZAMS} primary masses and orbital periods of the  \ac{S16} NSBH progenitors is shown in Figure~\ref{fig:S16_grids}.
We only show the  \ac{S16} population here because the \ac{F12d} population has far fewer NSBH binaries and their progenitors fall in similar regions on the \ac{HMS--HMS} grids as the  \ac{S16} NSBH binaries.

The different grid symbols, or ``termination flags", summarize the evolution of each of the pre-computed \texttt{MESA} models.
The cyan star and blue triangle markers plotted on top of the grids represent the NSBH progenitors in our  \ac{S16} \ac{BPS} population and distinguish between NSBHs that do and do not merge in a Hubble time, respectively.
NSBH binaries with initial mass ratios $\leq$~0.95 are potted on the $q=0.9$ grid slice, while those with mass ratios $>$~0.95 are plotted on the $q=1.0$ grid slice.
This division exists for plotting purposes only, as \texttt{POSYDON} interpolates discrete grid values to get the \ac{MT} class and evolution of each binary with its unique $q$, $P_{\mathrm{orb}}$, and $M_1$.

Looking at Figure~\ref{fig:S16_grids}, we see that all NSBH progenitors undergo \ac{MT} before  the first \ac{SN} occurs.
The termination flags with diamond markers represent models that ended their \ac{HMS--HMS} evolution by triggering a \ac{CE} event, while square markers indicate that one of the stars completed its evolution by reaching core carbon depletion.
The marker colors differentiate the evolutionary state of the donor star when the latest \ac{MT} phase was initiated, ranging from \ac{MS} (blue) to post-\ac{MS} (tan) to stripped \ac{He}--\ac{MS} (brown).
The green markers indicate that reverse \ac{MT} occurred (Section~\ref{sec:mass_transfer}).
Orange markers indicate that the binary underwent a contact phase where both stars filled their Roche lobes simultaneously at some point during the \ac{HMS--HMS} evolution.
The orange diamond markers specifically indicate that the binary terminated its \ac{HMS--HMS} evolution by triggering a double \ac{CE} event.

Many of the NSBHs on the $q=1.0$ grid fall near the boundary between the reverse \ac{MT} (green) and contact (orange) \ac{MT} classes. 
The most important consequence of this is that a given binary's probability of entering (double) \ac{CE} before the first \ac{SN} relies heavily on \texttt{POSYDON}'s classification of its \ac{HMS--HMS} \ac{MT} class around the class boundary, and thus it may be somewhat uncertain.
However, because we find that all NSBH progenitors experience a mass-ratio reversal before the first \ac{SN} regardless of the specific type of \ac{MT} that occurred, we do not expect this uncertainty to greatly impact our results apart from the prevalence of double \ac{CE} NSBH formation,  which impacts the NSBH orbital period-eccentricity distribution at \ac{DCO} formation (Section~\ref{sec:dco}) and the overall NSBH birth rate (Section~\ref{sec:alpha}).

We see that most binaries on the $q=0.9$ grid fall in the stable regions of evolution (square markers) and do not enter \ac{CE} before the first \ac{SN}, but many on the $q=1.0$ grid fall in regions that enter a double \ac{CE} (diamond markers) before the first \ac{SN}.
This difference in evolution distinguishes the Channel~I and Channel~II formation pathways for NSBHs discussed in Sections~\ref{sec:channel_I}~\&~\ref{sec:channel_II}.
We find that all of the NSBH progenitors on the $q=1.0$ grid with \ac{ZAMS} orbital periods $\gtrsim$ 40 days and primary mass $\gtrsim$ 25~$M_{\odot}$ (upper right region of NSBHs) enter a double \ac{CE} after the \ac{HMS--HMS} step, following the Channel II path of evolution.
The majority ($\sim$ 71\%) of these binaries merge in a Hubble time (cyan markers).
This is expected, as the double \ac{CE} phase significantly tightens the binary orbit and facilitates merging on shorter timescales (Section~\ref{sec:channel_II}).

\subsubsection{Channel I: Fully Detached Evolution}\label{sec:channel_I}

In Channel~I for forming NSBHs, the binary evolution is fully detached after the \ac{HMS--HMS} phase.
We find that 100\% of the NSBHs in the \ac{F12d} population and 36\% of the NSBHs in the \ac{S16} population form through this channel.
No \ac{MT} occurs in these binaries after the first \ac{SN}.
In particular, there is no Case BB \ac{MT} (where the donor is undergoing \ac{He}-shell burning during \ac{RLO}), which is thought to be necessary to spin-up pulsars with future \ac{NS} or white dwarf companions to \ac{MSP} speeds~\citep{tauris_formation_2012, lazarus_timing_2014, tauris_formation_2017}.
Channel~I NSBHs also do not experience any \ac{CE} evolution after the first \ac{SN}, so there is no possibility for unstable \ac{MT} onto the \ac{NS} to occur during \ac{CE}.

The lack of stable \ac{MT} arises from the fact that the \ac{He}-burning stellar progenitors of \acp{BH} in our NSBH populations are too heavy to expand and fill their Roche lobes.
All of their \ac{He} core masses before the second \ac{SN} are $>$~4~$M_{\odot}$, and many are much heavier, with masses of 10--11~$M_{\odot}$.
\ac{He}-burning stars with core masses $\gtrsim$ 4~$M_{\odot}$ can have radii up to two orders of magnitude smaller than \ac{He}-burning stars with lighter core masses~\citep{habets_evolution_1986}.
In addition, a \ac{CE} phase traditionally helps to shrink the binary orbit, making it easier for these heavy stars to initiate \ac{RLO}, but \ac{CE} is not triggered after the first \ac{SN} in these binaries (for more discussion on why \ac{CE} does not occur, refer to Section~\ref{sec:study_comparison}).

We find that 99\% of NSBH progenitors in our \ac{S16} population and 81\% of NSBH progenitors in our \ac{F12d} population have companion stars on the \ac{He}--\ac{MS} or post \ac{He}--\ac{MS} immediately following the \ac{HMS--HMS} stage of evolution when the primary has formed a \ac{NS}.
The remaining NSBH progenitors have companions that are \ac{H}-rich stars undergoing \ac{H}-core or \ac{H}-shell burning.
The binaries with \ac{H}-core burning companions have wide orbital periods ($>$ 150 days) that prevent them from undergoing \ac{RLO}.
The binaries with \ac{H}-shell burning companions have shorter orbital periods ($<$ 20 days), but most of them already underwent stable reverse \ac{MT} (where the secondary transfers mass to the primary) during the \ac{HMS--HMS} evolution.
Thus, \ac{RLO} from the secondary companion already occurred before the \ac{NS} formation, preventing accretion onto the \ac{NS}.

The breakdown of how many NSBHs formed through Channel~I that do and do not merge in a Hubble time is shown in Figure~\ref{fig:form_channels}.
We see that most binaries remain detached, which is due to the fact that there is no mechanism such as (double) \ac{CE} to help tighten the binary orbit.
We find that only 16\% of NSBHs in the \ac{F12d} population and 12\% of NSBHs in the \ac{S16} population merge in a Hubble time if formed through Channel I.
Most of the binaries that do merge have very high eccentricities ($>$ 0.9) upon \ac{DCO} formation (Figure~\ref{fig:NSBH_porb_e}), which allows them to merge on shorter timescales~\citep[][]{peters_gravitational_1964}.

\subsubsection{Channel II: Double \ac{CE} Evolution}\label{sec:channel_II}

In Channel~II, NSBH progenitors undergo a double \ac{CE} after the \ac{HMS--HMS} phase of evolution and before the first \ac{SN}.
Channel~II dominates the NSBH formation in the  \ac{S16} population, accounting for 64\% of its NSBHs. 
No NSBHs form through this channel in the \ac{F12d} population.
This is because these binaries have primary masses $\lesssim$ 20~$M_{\odot}$ at \ac{ZAMS}, and thus they are not massive enough to fall inside the region of the \ac{HMS--HMS} grids where binaries undergo a double \ac{CE}.
The double \ac{CE} region is located at $P_{\mathrm{orb}}$ $\gtrsim$ 40 days and $M_1$ $\gtrsim$ 25~$M_{\odot}$ in the $q=1.0$ grid of Figure~\ref{fig:S16_grids}.

The main difference between the NSBH populations formed through this channel vs. Channel~I is that there is a higher fraction of merging NSBHs in Channel~II.
About 71\% of NSBHs formed through Channel~II in the  \ac{S16} population merge in a Hubble time (Figure~\ref{fig:form_channels}).
This is because the double \ac{CE} phase tightens the binary orbit, which greatly facilitates the ability to merge on shorter timescales.
We find that all NSBHs formed through Channel~II have orbital periods $<$ 0.5 days after the double \ac{CE} phase, even though 87\% of them start with orbital periods $>$ 100 days before the double \ac{CE} begins.
The reason that these binaries go through a double \ac{CE} rather than a single \ac{CE} before the first \ac{SN} is that they contain stars of near-equal mass at \ac{ZAMS} (Figure~\ref{fig:mass_ratios}).
Their stellar components evolve on similar timescales, which causes them to reach post--MS stages of evolution with expanded envelopes at around the same time.
In \texttt{POSYDON}, if unstable \ac{MT} is triggered and both stars are in \ac{RLO} with a giant-like structure and core-envelope separation, the binary undergoes a double \ac{CE} (Section ~\ref{sec:CE}).

After the first \ac{SN} occurs and the \ac{NS} is formed, the evolution for these binaries proceeds identically to those formed through Channel I, i.e. the evolution is fully detached and there is no \ac{MT} onto the \ac{NS} before the \ac{BH} is formed.

\subsection{NSBH DCO Properties}\label{sec:dco}

\begin{figure}
\includegraphics[width=0.5\textwidth]{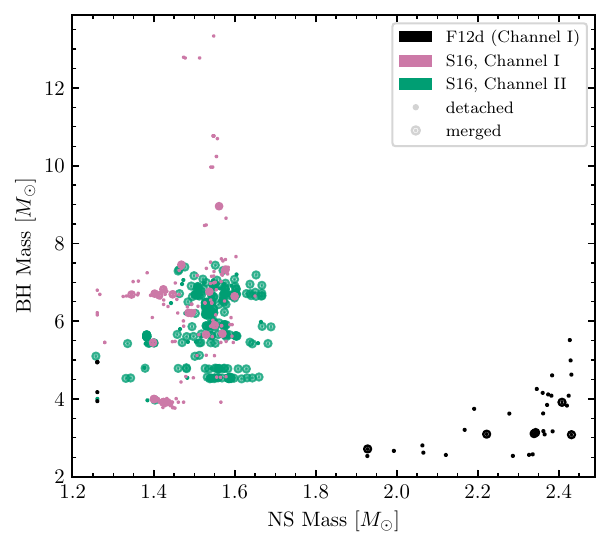}
\caption{
\ac{BH} vs. \ac{NS} masses for all NSBH binaries at their formation in both the \ac{F12d} (black) and  \ac{S16} (green/pink) populations. 
Each population is separated into binaries that merge and do not merge in a Hubble time with the large and small markers, respectively.
For the \ac{S16} population, the NSBHs are also separated by Channel~I vs. Channel~II formation (Sections~\ref{sec:channel_I} \& \ref{sec:channel_II}).
All \ac{F12d} NSBHs form through Channel I.
}
\label{fig:NSBH_masses}
\end{figure}

\begin{figure}
\includegraphics[width=0.5\textwidth]{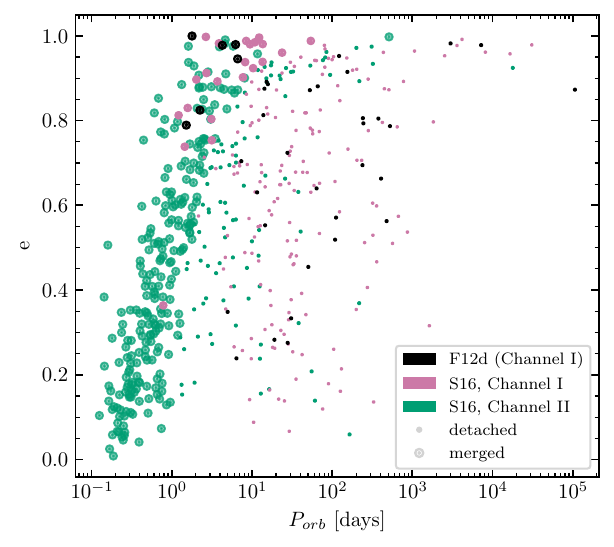}
\caption{Eccentricity vs. orbital period for all NSBH binaries at their formation in both the \ac{F12d} (black) and  \ac{S16} (green/pink) populations. 
Each population is separated into binaries that merge and do not merge in a Hubble time with the large and small markers, respectively.
For the \ac{S16} population, the NSBHs are also separated by Channel~I vs. Channel~II formation (Sections~\ref{sec:channel_I} \& \ref{sec:channel_II}).
All \ac{F12d} NSBHs form through Channel I.
}
\label{fig:NSBH_porb_e}
\end{figure}

We plot the component masses at \ac{DCO} formation for all NSBHs in our populations in Figure~\ref{fig:NSBH_masses}. 
The green and pink points represent the  \ac{S16} population and the black points represent the \ac{F12d} population.
For the \ac{S16} population, the NSBHs are separated by their formation channels, with the pink points representing Channel~I (Section~\ref{sec:channel_I}) and the green points representing Channel II (Section~\ref{sec:channel_II}).
The \ac{F12d} NSBHs do not need to be separated because all of them form through Channel I.
The larger markers designate binaries that merge in a Hubble time, while the smaller markers represent binaries that remain detached.

The most notable difference between the \ac{F12d} and  \ac{S16} NSBHs is the dominance of high-mass ($>$ 2~$M_{\odot}$) \acp{NS} in the \ac{F12d} population.
The \ac{F12d} \ac{SN} prescription will form \acp{NS} with masses up to the designated maximum \ac{NS} mass limit~\citep{fryer_compact_2012}, which we set to 2.5~$M_{\odot}$, whereas the  \ac{S16} prescription does not form \ac{NS} above about 1.8~$M_{\odot}$~\citep{sukhbold_core-collapse_2016}. 
Most of the \ac{F12d} binaries with high-mass \acp{NS} have very low-mass \ac{BH} companions ($<$ 3~$M_{\odot}$), implying their mass ratios remain close to unity even after both \ac{CO}s have formed.
This is a consequence of the \ac{F12d} \ac{SN} mechanism, which produces more high-mass \ac{NS} and low-mass \acp{BH} for NSBHs in general (see discussion in Section~\ref{sec:frequencies}). 
However, we note that so far in the literature, the only explanation for the presence of low-mass BHs in \ac{GW} sources and the existence of a low-end mass gap in X-ray binaries indicates that \ac{NS} masses at birth may be limited to $<$ 2~$M_\odot$ like in the \ac{S16} prescription~\citep{siegel_investigating_2023}.

There are a few low-mass \acp{NS} in both the \ac{F12d} and \ac{S16} populations that have masses of 1.26~$M_{\odot}$.
All of these \acp{NS} formed through \acp{ECSN}.
In \texttt{POSYDON}, the proto-\ac{NS} mass is set uniformly for all stars undergoing \acp{ECSN} with the ~\citet{podsiadlowski_effects_2004} prescription.
The difference in modeling of \acp{ECSN} vs. \acp{CCSN} forms a gap in the \ac{NS} masses for the \ac{F12d} NSBHs.

The \ac{S16} NSBH binaries have heavier \ac{BH} components and a wider distribution of \ac{DCO} mass ratios compared to the \ac{F12d} binaries.
As discussed in Section~\ref{sec:frequencies}, this is because the type of \ac{CO} formed from a given \ac{SN} with the  \ac{S16} mechanism is more stochastic than the \ac{F12d} mechanism.
Thus, even though these binaries may have \ac{ZAMS} mass ratios close to one, this does not guarantee that the final \ac{DCO} binary will also have a similar mass ratio.
In addition, the \ac{S16} \ac{SN} prescription does not produce \acp{BH} $\lesssim$ 3.7~$M_{\odot}$ or \acp{NS} $\gtrsim$ 1.8~$M_{\odot}$ with the N20 engine~\citep{sukhbold_core-collapse_2016}, meaning there is a small gap between the heaviest \acp{NS} and lightest \acp{BH} that are formed.

We observe that the \ac{S16} NSBHs with the most unequal component masses (\acp{BH} $>$ 8~$M_{\odot}$) all form through Channel~I.
The Channel~I systems can have more unequal \ac{ZAMS} masses compared to the Channel~II systems, which must have \ac{ZAMS} mass ratios $>$ 0.95 in order to initiate the double \ac{CE} phase (Sections~\ref{sec:hms_hms} \& \ref{sec:channel_II}).

In Figure~\ref{fig:NSBH_porb_e}, we plot the eccentricity vs. orbital period of all NSBHs in our populations.
Similar to Figure~\ref{fig:NSBH_masses}, the green and pink points represent the \ac{S16} population and the black points represent the \ac{F12d} population, while the two colors for the \ac{S16} NSBHs differentiate between formation channels and the marker sizes distinguish between binaries that do and do not merge in a Hubble time.

For both the \ac{F12d} and \ac{S16} populations, the merging NSBHs have shorter orbital periods than the detached NSBHs at \ac{DCO} formation.
This is especially true for the \ac{S16} Channel II systems, as they experience a double \ac{CE} phase that significantly tightens the binary orbit before the first \ac{SN}.
Most of the mergers formed through Channel~I in both the \ac{S16} and \ac{F12d} populations have very high eccentricities with slightly greater orbital periods; the high eccentricity is necessary to merge these binaries in a Hubble time.

\subsection{Model Variations}
\subsubsection{Reduced \ac{CCSN} Kicks}
As mentioned in Section~\ref{sec:supernovae}, we analyze the effects of reducing \ac{CCSN} natal kicks by drawing kicks from a Maxwellian using the reduced kick scatter estimates from~\citet{mandel_simple_2020}.
In this study, they predict a natal kick scatter of 120~km~s$^{-1}$ for \acp{NS} and 60~km~s$^{-1}$ for \acp{BH} formed via \ac{CCSN}, which we set as the respective velocity dispersions for our Maxwellian kick distributions.
These velocity dispersion estimates are greatly reduced compared to the scatter of 265~km~s$^{-1}$ for all \acp{NS} and \acp{BH} in our default models~\citep{hobbs_statistical_2005}.
Our adoption of the kick scatter estimates from \citet{mandel_simple_2020} does not consider their full kick model, which draws kicks from a Gaussian distribution centered at the normalized kick scatter values rather than a Maxwellian distribution.

In principle, we would expect reduced \ac{SN} kicks to increase the number of bound \ac{NS}--\ac{BH} in our populations, as these systems are less likely to be disrupted upon \ac{CO} formation.
We find that this is indeed the case.
Looking at Table~\ref{table:frequencies}, we see that the NSBH birth rate is an order of magnitude higher for the \ac{F12d} population and over twice as high for the  \ac{S16} population with the reduced kicks (designated as ``low $\sigma_{\mathrm{CCSN}}$").
BHNSs still dominate the total \ac{NS}--\ac{BH} populations, and the birth rate for these systems is about four times higher for the \ac{F12d} population and three times higher for the  \ac{S16} population.

Qualitatively, all of the remaining results discussed in Section \ref{sec:results} are comparable between the NSBH populations formed with reduced \ac{CCSN} kicks and our default kick settings.
The primary mass distributions at \ac{ZAMS} and the mass ratio distributions at \ac{ZAMS} do not change for either the \ac{F12d} or the  \ac{S16} populations.
The formation channels of NSBHs are the same, and the ratios of merged and detached binaries formed through Channel~I and Channel~II are comparable as well.
Lastly, the qualitative NSBH properties at \ac{DCO} formation exhibit no significant changes between the model variation and the default.

\subsubsection{Reduced \ac{CE} Efficiency}\label{sec:alpha}
The \ac{CE} efficiency parameter $\alpha$ (Section~\ref{sec:CE}) impacts the ``success" of the double \ac{CE} events for NSBHs formed through Channel II.
We set $\alpha$ = 1 in our default models, which is a relatively optimistic setting for allowing envelopes to be successfully ejected during \ac{CE}.
To evaluate the impact of $\alpha$ on our results, we run two model variations with $\alpha$~=~0.5 and $\alpha$~=~0.1, which we expect to make it more difficult for binaries to survive \ac{CE}.
We only run these model variations with the \ac{S16} remnant mass prescription because NSBHs formed through Channel II do not exist in the \ac{F12d} population.

We find that changing the value of $\alpha$ does not introduce any new possible formation channels for NSBHs, but lowering the value of $\alpha$ ``kills" most of the Channel II systems, which in turn lowers all \ac{NS}--\ac{BH} birth rates.
We display the \ac{S16} birth rates with $\alpha$~=~0.5 and $\alpha$~=~0.1 in Table~\ref{table:frequencies}.
In particular, the NSBH birth rate drops by about 62\% when lowering $\alpha$ from 1 to 0.5.
As $\alpha$ decreases, increasingly more potential NSBH progenitors do not survive the double \ac{CE} phase and instead merge before the \ac{NS} formation.
We find that only 13\% of NSBHs form through Channel II with $\alpha$~=~0.5 and no binaries form through Channel II with $\alpha$~=~0.1 (as opposed to 64\% with $\alpha$~=~1).
All of the $\alpha$~=~0.5 NSBHs that do form through Channel II merge in a Hubble time.
This is expected, as decreasing $\alpha$ also shrinks the separation of the binary after the envelope ejection~\citep[e.g.,][]{mapelli_cosmic_2018}.

The Channel~I NSBH results do not change as $\alpha$ is decreased, so the low-$\alpha$ \ac{S16} populations look approximately identical to Figures~\ref{fig:form_channels},~\ref{fig:NSBH_masses}, and~\ref{fig:NSBH_porb_e} without the Channel~II population present.

\section{Discussion}\label{sec:discussion}

\subsection{Comparison with Previous Studies}\label{sec:study_comparison}
We present these results in conversation with previous studies that model \ac{BPS} populations of NSBHs in the Milky Way, particularly as they relate to potential \ac{MSP}--\ac{BH} binaries. 
A recent study by \citet{chattopadhyay_modelling_2021} examines Galactic NS--BH populations using the \texttt{COMPAS} code~\citep{stevenson_formation_2017, vigna-gomez_formation_2018, neijssel_effect_2019}.
In contrast to our results, they find that \texttt{COMPAS} can produce populations of NSBHs with recycled \acp{NS}, allowing for the presence of a detectable Galactic \ac{MSP}--BH population.
The dominant formation channel for these systems is as follows: the initially more massive primary initiates stable \ac{MT} onto the secondary during the \ac{HMS--HMS} evolution and then eventually explodes to form a \ac{NS}.
The secondary then also expands to fill its Roche lobe, which initiates unstable \ac{MT} and a \ac{CE} phase that facilitates significant accretion onto the \ac{NS}.
If the \ac{CE} is successful, the binary orbit dramatically shrinks and an instance of stable Case BB \ac{MT} phase may occur before the second \ac{SN}, depending on the mass of the secondary star.
These \ac{MT} phases after the first \ac{SN}, particularly the \ac{MT} during \ac{CE}, are required to obtain \ac{MSP} populations in \texttt{COMPAS} that are consistent with Galactic double \ac{NS} observations~\citep{chattopadhyay_modelling_2021}.

In \texttt{COMPAS}, \ac{CE} initiation after the \ac{NS} formation is possible due to their treatment of stable \ac{MT}, which leads to  the formation of \ac{NS}--\ac{BH} binaries with asymmetric masses in their populations ~\citep{broekgaarden_impact_2021}.
Their \ac{MT} prescription allows for more efficient accretion during the \ac{HMS--HMS} evolution than the \ac{MT} treatment in \texttt{POSYDON}'s pre-computed \texttt{MESA} models, as stellar accretion in \texttt{COMPAS} is only constrained by the thermal timescales of the donor and the accretor and does not consider the accretor's rotation rate~\citep{stevenson_formation_2017}.
As a result, \citet{chattopadhyay_modelling_2021} find that the first episode of \ac{MT} is almost completely conservative for \ac{NS}--\ac{BH} progenitors, which in their model means that all the mass lost by the donor is successfully accreted by the companion. 
This implies that mass ratio reversal can happen for more asymmetric \ac{ZAMS} masses.
Because the progenitor stars are evolving on different timescales in this case, a single \ac{CE} after the first \ac{SN} is allowed to take place.

In contrast, mass accretion onto non-degenerate companions in \texttt{POSYDON} is limited by the critical rotation rate of the accretor, as discussed in Section~\ref{sec:mass_transfer}~\citep{fragos_posydon_2023}.
This means that the actual amount of mass accreted by the secondary may be very low depending on its angular momentum and stellar structure, even when assuming initially conservative \ac{MT} between the donor and accretor.
Such a scenario is not uncommon, as non-degenerate accretors will gain angular momentum and can often reach critical rotation rates during \ac{MT}.
The result of this is that mass ratio reversal during \ac{HMS--HMS} evolution only occurs for binaries with \ac{ZAMS} mass ratios near unity, and potential NSBH progenitors with more unequal masses are disrupted (Section~\ref{sec:form_channels}).

It is important to note that \citet{chattopadhyay_modelling_2021} find infrequent instances of NSBH formation nearly identical to our Channel II binaries.
For \texttt{COMPAS} \ac{NS}--\ac{BH} binaries with \ac{ZAMS} mass ratios very close to one, the stars evolve on similar timescales, which causes them to enter a double \ac{CE} phase before the first \ac{SN}.
Because the \ac{NS} now forms after the \ac{CE}, it does not undergo any mass accretion and thus \ac{BH} binaries with recycled \acp{NS} do not form.
In this case, our results are in agreement with one another.

The study done by \citet{kruckow_progenitors_2018} examines Galactic \ac{DCO} populations with the \texttt{ComBinE} code~\citep{tauris_origin_1996, voss_galactic_2003}.
When looking at NSBH formation channels, they find that no \ac{CE} or stable \ac{MT} occurs after the first \ac{SN}, and thus recycled \acp{NS} with \ac{BH} companions do not form at all in their populations.
They do not treat double \ac{CE} evolution, so all of their NSBHs form through our Channel~I evolution.
The key to this result is that they assume highly inefficient \ac{MT}, which, like in our study, requires their NSBH progenitors to have near-equal \ac{ZAMS} masses for mass ratio reversal to occur during the \ac{HMS--HMS} evolution~\citep{kruckow_progenitors_2018}.
In this sense, our results are in agreement with one another.
However, while they make the inefficient \ac{MT} assumption in efforts to better match their populations to double \ac{NS} observations, it is important to note that their findings are sensitive to this fine-tuned setting of \ac{MT} parameters, while in \texttt{POSYDON} the \ac{MT} is more robustly and self-consistently handled by the pre-computed \texttt{MESA} models.

\subsection{Model Uncertainties}
As discussed throughout Section~\ref{sec:results}, our results are most impacted by the \ac{CC} prescription used in our \ac{BPS} models.
Changing this prescription between \ac{F12d} and  \ac{S16} impacts the \ac{NS}--\ac{BH} birth rate by multiple orders of magnitude (Table~\ref{table:frequencies}).
In addition, it determines the possible NSBH formation channels (Section~\ref{sec:channel_II}), which not only controls the (absence of) recycled \acp{NS}, but also affects the NSBH \ac{DCO} properties (Figures ~\ref{fig:NSBH_masses}~\&~\ref{fig:NSBH_porb_e}).
\citet{chattopadhyay_modelling_2021} also find that the birth rates and properties of \ac{NS}--\acp{BH} change significantly when comparing different \ac{SN} remnant mass prescriptions.
For example, they report the total \ac{NS}--\ac{BH} birth rate is over twice as high for populations run with the Fryer--12 rapid prescription compared to the Fryer--12 delayed prescription, and the \ac{ZAMS} mass ratios of the rapid NSBHs are significantly lower than those of the delayed NSBHs.
This further confirms that the formation rates, channels, and properties of NSBHs are very sensitive to how remnant masses and remnant types are determined, which makes sense given that these systems require relatively specific pre-\ac{SN} conditions to form successfully.
Thus, future \ac{CC} prescriptions could potentially introduce new NSBH formation channels not identified in this paper and change the NSBH population properties.

We also reiterate the model uncertainties related to double \ac{CE} evolution discussed in Section~\ref{sec:CE}.
The physical conditions that lead to the onset of a double \ac{CE} are thought to differ from those for a single \ac{CE} because \ac{MT} is not expected to become unstable in the same manner for post-\ac{MS} stars of near-equal mass~\citep{ivanova_common_2020}.
In \texttt{POSYDON}, we assume that all binaries with two components in \ac{RLO} that contain at least one post-\ac{MS} stellar component immediately initiate unstable \ac{MT}, and this condition in particular could lead to an over-estimation of the number of binaries undergoing double \ac{CE} in our populations.
The amount of double \ac{CE} NSBHs is also sensitive to the \ac{MT} classification performed by \texttt{POSYDON}'s initial-final interpolation scheme, as most of the double \ac{CE} systems lie on a classification boundary in the \ac{HMS--HMS} grids (Section~\ref{sec:hms_hms}).
Improvements in the classification method could change how many NSBHs go through a double \ac{CE} phase.

NSBH formation is impacted by the \ac{He}-burning stars that serve as \ac{NS} companions before the second \ac{SN} (Section~\ref{sec:channel_I}).
The behavior of these stars in relation to the binary orbit primarily determines if \ac{MT} onto the \ac{NS} will occur or not.
Changes in the related input physics of the pre-computed \texttt{MESA} models, such as the stellar wind scheme used for these \ac{He} stars, could potentially affect our results.
Changing stellar winds would not only impact the total amount of stellar mass-loss, but could also change the orbital separation of the binary.
Currently, stellar winds for high-mass stars ($>$ 8 $M_{\odot}$) in \texttt{POSYDON} are treated using the \texttt{MESA Dutch} scheme, which captures key features of massive stellar evolution~\citep{fragos_posydon_2023}.

In addition, the method of assigning binary properties from the \texttt{MESA} grids during \ac{BPS} can impact the resulting populations.
We use \texttt{POSYDON}'s initial-final interpolation scheme in all of our binary populations (Section~\ref{sec:methods}).
To assess potential uncertainties due to this choice of interpolator, we also run populations using \texttt{POSYDON}'s nearest-neighbor scheme for acquiring binary properties ~\citep{fragos_posydon_2023, andrews_posydon_2024}.
We find that our qualitative results, such as the distinct NSBH formation channels, do not change when assigning binary properties with nearest-neighbor instead of initial-final interpolation.
We find that our quantitative results with nearest-neighbor, such as the birth rate of NSBHs, scale in a similar manner as when lowering the \ac{SN} kicks.

Lastly, our results are sensitive to the resolution of \texttt{POSYDON}'s pre-computed grids as well as the presence of unconverged \texttt{MESA} models in these grids.
An increased grid resolution will improve the accuracy of the binary \ac{MT} classification and the interpolation of binary properties.

\subsection{Implications for MSP--BH Detections}

As discussed in Section~\ref{sec:intro}, only one possible \ac{MSP}--\ac{BH} system has been detected thus far; the companion object falls in the lower mass gap and cannot be differentiated between a low-mass \ac{BH} and high-mass \ac{NS}~\citep{barr_pulsar_2024}. 
Because of its location in the globular cluster NGC 1851, it is predicted that this system formed dynamically, where the \ac{MSP} was first spun--up by a low-mass companion that was later replaced by a high-mass companion through exchange interactions~\citep{barr_pulsar_2024}.
Theoretical models predict that it is relatively difficult to form \ac{NS}--\ac{BH} systems in \acp{GC}~\citep[e.g.,][]{clausen_dynamically_2014, ye_rate_2019, arca_sedda_dissecting_2020, hoang_neutron_2020}, but \ac{MSP} formation is thought to be more efficient in these environments~\citep[e.g.,][]{manchester_australia_2005, ransom_pulsars_2008, bahramian_slar_2013}.
\citet{ye_millisecond_2019} find that the number of \acp{MSP} in a given cluster may even be anti-correlated with the number of \acp{BH}, making \ac{MSP}--\ac{BH} formation unlikely.

Other potential formation environments could include \ac{AGN} such as the Galactic center~\citep[e.g.,][]{fragione_black_2019-1, stephan_fate_2019, mckernan_black_2020} or isolated stellar triples~\citep[e.g.,][]{toonen_evolution_2016, liu_black_2018, fragione_black_2019}, both of which allow for more complex channels of \ac{NS}--\ac{BH} formation.
However, the formation of \ac{MSP}--\ac{BH} binaries in these environments has not been studied in detail.

Past \ac{BPS} studies have examined the formation of isolated NSBHs in Milky Way-like galaxies, but come to opposing conclusions regarding the ability of these systems to recycle \acp{NS}~\citep{kruckow_progenitors_2018, chattopadhyay_modelling_2021}.
We find that while NSBHs can form in the Galactic field, the pulsars in these binaries will not be recycled.

With all of this in consideration, our results have important implications for current and future pulsar surveys targeting pulsar--\ac{BH} systems.
If surveys wish to observe a \ac{MSP}--\ac{BH} system specifically, it is more likely that this will happen by targeting dynamically active environments (such as Galactic \acp{GC} or nuclear clusters).
However, we cannot assess the uncertainties related to NSBH formation or observations in these environments in the current study.

\section{Conclusions}\label{sec:conclusion}
We simulate populations of binaries at solar metallicity using the \texttt{POSYDON} population synthesis code and examine the properties of the \ac{NS}--\ac{BH} subpopulation, including both merged and detached systems.
We find that \ac{NS}--\ac{BH} binaries in which the \ac{NS} forms first (NSBHs) are significantly less common than those where the \ac{BH} forms first (BHNSs).
The Galactic birth rate of NSBHs is highly dependent on the \ac{CC} prescription, kick prescription, and \ac{CE} efficiency parameter $\alpha$ used in modeling the binary population.
For example, we find that the birth rate of NSBHs is about an order of magnitude higher with the  \ac{S16} remnant mass prescription compared to the \ac{F12d} prescription, and these differences are even greater when lowering \ac{CCSN} kicks but diminish when reducing the value of $\alpha$.

All NSBHs in our populations have \ac{ZAMS} mass ratios close to unity, and all of them undergo a mass ratio reversal before the first \ac{SN} via \ac{MT} during \ac{HMS--HMS} evolution.
The mass accretion of the secondary during \ac{HMS--HMS} evolution is relatively inefficient due to \texttt{POSYDON}'s treatment of rotationally-limited accretion and boosted stellar winds.
After the first \ac{SN}, NSBHs form through two different channels.
In Channel I, the binary undergoes fully detached evolution after the \ac{HMS--HMS} phase, all the way through the second \ac{SN}.
In Channel II, the binary experiences a double \ac{CE} phase of evolution after the \ac{HMS--HMS} phase and before the first \ac{SN}.
In both cases, there is no \ac{MT}, stable or unstable, in the binary after the first \ac{SN}.

The lack of \ac{MT} after the first \ac{SN} implies that there is no possibility of accretion onto the \ac{NS} in our NSBH binaries.
Accretion is required to recycle the \ac{NS}, and thus we conclude that \ac{MSP}--{BH} binaries cannot be formed.
We suggest that ongoing pulsar surveys are more likely to detect these systems in dynamically active environments in the Milky Way.

\section*{Acknowledgements}

The authors thank Debatri Chattopadhyay and Chase Kimball for their input on our methods and results.  
C.L.\ and V.K.\ acknowledge support from the Gordon and Betty Moore Foundation (grant awards GBMF8477 and GBMF12341), CIERA, and Northwestern University. V.K.\ was partially supported through the D.I.Linzer Distinguished University Professorship fund.
P.M.S, E.T, K.A.R., M.S., and S.G. were also supported by the project numbers GBMF8477 and GBMF12341.
    J.J.A.~acknowledges support for program number JWST-AR-04369.001-A provided through a grant from the STScI under NASA contract NAS5-03127. 
    S.S.B., T.F., M.U.K., and Z.X. were supported by the the Swiss National Science Foundation, project number PP00P2\_211006. 
    S.S.B. and M.M.B. were supported by the the Swiss National Science Foundation, project number CRSII5\_213497. 
    M.M.B. is also supported by the Boninchi Foundation and the Swiss Government Excellence Scholarship. 
    K.K. is supported by a fellowship program at the Institute of Space Sciences (ICE-CSIC) funded by the program Unidad de Excelencia Mar\'ia de Maeztu CEX2020-001058-M. 
    Z.X. acknowledges support from the Chinese Scholarship Council (CSC). 
    E.Z. acknowledges support from the Hellenic Foundation for Research and Innovation (H.F.R.I.) under the “3rd Call for H.F.R.I. Research Projects to support Post-Doctoral Researchers” (Project No: 7933). 
    The computations were performed at Northwestern University on the Trident computer cluster (funded by the GBMF8477 award) and at the University of Geneva on the Yggdrasil computer cluster. 
    This research was partly supported by the computational resources and staff contributions provided for the Quest high-performance computing facility at Northwestern University, jointly supported by the Office of the Provost, the Office for Research and Northwestern University Information Technology.

\textit{Software:} \texttt{POSYDON}~\citep{fragos_posydon_2023, andrews_posydon_2024}; \texttt{Matplotlib}~\citep{hunter_matplotlib_2007}; \texttt{NumPy}~\citep{van_der_walt_numpy_2011}; \texttt{Pandas}~\citep{mckinney_data_2010}.

\bibliography{references_manual}{}
\bibliographystyle{aasjournal}

\end{document}